# Foundation Models for Geophysics: Reviews and Perspectives

Qi Liu,  Jianwei Ma*

*School of Earth and Space Sciences, Institute for Artificial Intelligence, Peking University, Beijing, China*

*Corresponding email: jwm@pku.edu.cn

**Abstract** Recently, large models, or foundation models have demonstrated outstanding performance and have been applied in a variety of disciplines, such as chemistry, biology, economics, etc. Foundation models, trained on vast amounts of data, can be adapted to a wide range of use cases. The emergence of foundation models has a significant impact on the research paradigms in these fields. Geophysics is a scientific field dedicated to exploring and understanding the Earth's structures and states through the application of physical principles and the analysis of multimodal geophysical data. In the field of geophysics, the processing and interpretation of geophysical data are characterized by three primary features: extensive data volume, multimodality, and dependence on experience. These characteristics provide a suitable environment as well as challenges for the development and breakthrough of foundation models in the field of geophysics. In this perspective, we discuss the potential applications and research directions of geophysical foundation models (GeoFMs), exploring new research paradigms in geophysics in the era of foundation models. Exploration geophysics is the main focus, while the development of foundation models in remote sensing, seismology, and other related sub-disciplines in geophysics is also discussed. In the meantime, we also propose two strategies for constructing GeoFMs and discuss challenges that may arise during the process of development.

# 1 Introduction

As artificial intelligence transitions from rule-based systems to the era of machine learning, data-driven deep learning methods have emerged as a key breakthrough, leveraging neural networks to learn various patterns from data for different tasks. With the rapid growth of data volumes and continuous advancements in computational resources, deep learning-based foundation models have increasingly come into focus. Foundation models are large-scale models trained on vast and diverse datasets that can be applied across a wide range of tasks and operations. In recent years, the emergence of foundation models has achieved great success in various fields, including natural language processing models represented by ChatGPT[1], image segmentation models such as Segment Anything Model[2] (SAM), and video processing models represented by Sora[3], among others. These deep learning-based foundation models are trained on massive amounts of data and large numbers of model parameters, showcasing the astounding upper limits of deep learning methods. The exceptional performance of foundation models has profoundly influenced the research paradigms in various fields, such as chemistry[4,5], biology[6], finance[7,8], medicine[9], and remote sensing[10-12], among others. However, the application and development of foundation models in geophysics are still in an initial stage.

Geophysics is a scientific discipline that investigates and analyzes the Earth's structures and states using physical principles and multimodal geophysical data. For instance, exploration geophysics utilizes geophysical techniques and seismic data to study the Earth's subsurface structure, aiding in the discovery of natural resources such as oil and minerals. Additionally, sub-disciplines like remote sensing and seismology offer further insights; remote sensing employs satellite images to study the states of the Earth's surface and to monitor environmental changes, while seismology focuses on the study of the earthquakes and delivers insights into the Earth's internal composition and dynamics.

These sub-disciplines of geophysics provide a comprehensive understanding of the Earth's structure and states. The procedure of geophysics mainly includes three stages: data acquisition (multimodal geophysical data, such as seismic data, remote sensing images, gravity data, atmosphere data, etc.), data processing (multiple tasks, such as first-arrival picking, interpolation, denoising), and data interpretation (seismic imaging, wheather forecasting, earthquake detection, and so on). The data processing stage has the following features: (1) large amounts of data; (2) multimodal data; (3) multiple tasks. The interpretation stage significantly relies on human analysis and experience. The extensive multimodal data serve as a basis for the training and fine-tuning of large models, while the multi-task and highly experience-dependent features offer a broad space for the applications of GeoFMs in the field of geophysics.

Over the past decade, the rapid development of deep learning methods has had a profound impact on the research paradigm in geophysics, which has shifted from traditional methods to deep learning-based data-driven approaches[13], and further to foundation models, as shown in Figure 1. To illustrate the paradigm shift, we use exploration geophysics as an example. Traditional methods are usually model-driven and follow specific assumptions. Moreover, they are intricately dependent on the experience of experts. For example, the linearity assumption presumes that the seismic events exhibit linearity in small windows[14], while transform-domain-based methods assume that the data are sparse[15] or low-rank after transformations[16], like the Fourier[17], Curvelet[18], and Radon[19] transforms. These traditional methods are effective in the seismic data processing stage. However, when faced with complex field data, they may not be able to obtain reasonable denoising or interpolation results due to the invalid assumptions. As the data volumes grow exponentially across various domains, data-driven deep learning methods have become one of the most focused research directions in all disciplines.

Deep learning methods have been utilized in various fields of exploration geophysics, such as first-arrival picking[20], interpolation[21], inversion[22], etc. Review articles about deep learning in geophysics have recently been published[13,23], providing a detailed account of the application of deep learning methods in geophysics. Nonetheless, the current deep learning methods in geophysics are mostly based on the limited parameter volumes and training data, which limits the generalization of the model and thereby restricts the application of deep learning-based methods in the exploration industry. In light the developments of foundation models, these limitations are gradually being alleviated. GeoFMs, trained on a large amount of multimodal geophysical data, demonstrate strong generalization capabilities across various geophysical downstream tasks, assisting researchers in conducting academic studies and exploration operations in the field of geophysics.

The success of foundation models across various fields has demonstrated the immense potential of deep learning methods, which are bound to bring profound changes into the paradigm of geophysical research. This article aims to provide a glance at the recent foundation models research related to geophysics, analyze the impact of the era of foundation models on the paradigm of geophysical research, and discuss the potential research directions for GeoFMs. We introduce the background, applications and development of foundation models in various disciplines, followed by a presentation of potential applications of GeoFMs in geophysics, including a first-arrival picking method using SAM, which demonstrates the substantial potential of foundation models in the field of geophysics. Then, we discuss the challenges faced by the development of GeoFMs in geophysics and possible research directions.

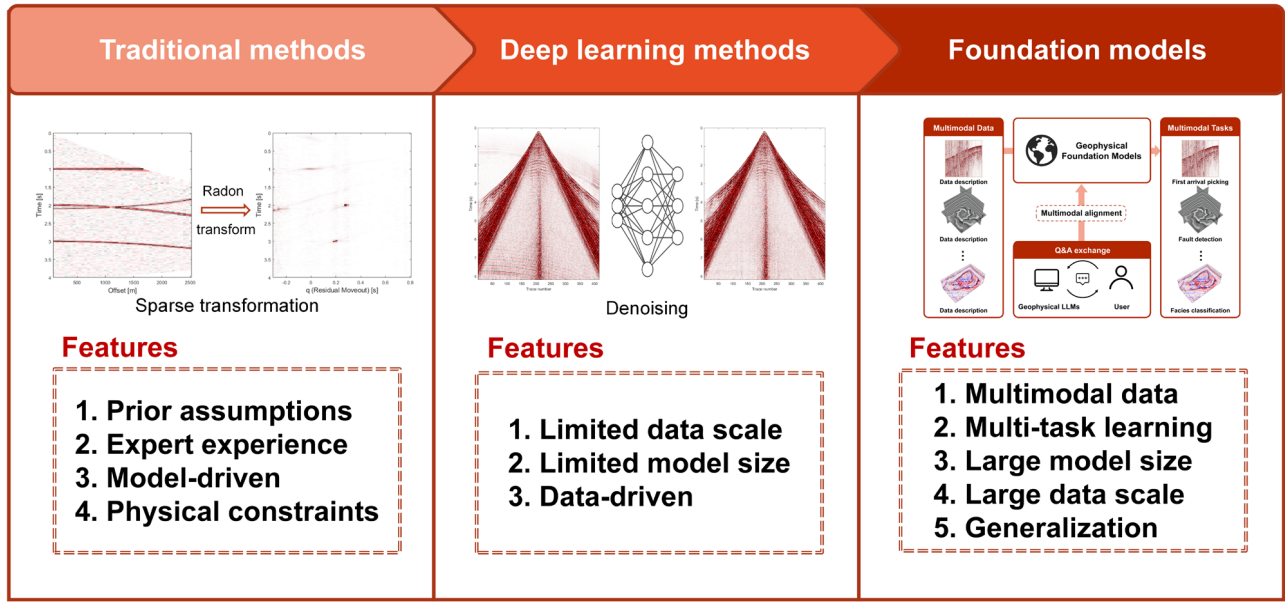

**Figure** 1 Research paradigm shift in geophysics, the transition from traditional methods to deep learning methods and foundation models with corresponding features.

## 2 The Background of Foundation Models

### 2.1 Large Language Models

Large Language Models (LLMs) are pre-trained on an immense amount of text data[24,25], and are capable of handling any text-formatted task without the necessity of task-specific training. The zero-shot generalization capability of LLMs is attributed to the in-context learning paradigm[26], which recognizes various patterns in natural language and learns prompts through next-token prediction autoregressive training.

The introduction of the Transformer framework[27] has propelled the rapid advancement of LLMs. The Transformer model, built upon the self-attention mechanism, is capable of capturing long-distance dependence in text and enables parallel training with large parameters, thereby laying the methodological foundation for LLMs training. LLMs require extensive text data collected from various sources and are trained on thousands of GPUs or TPUs, with the number of parameters reaching the order of billions. Here, we briefly introduce several representative LLMs. The GPT series[1], proposed by OpenAI, uses a Transformer Decoder-only architecture and leverages reinforcement learning from human feedback (RLHF)[28] technology. It is currently the most influential LLM. Google's PaLM series[29] has achieved efficient training of super-large parameters on TPUs with breakthrough performance in many downstream tasks. Meta AI released the LLaMA series[30] and open-sourced the model parameters, providing a foundation for the research of LLMs. Recently, the Claude 3 series[31] launched by Anthropic has made breakthrough progress in text tasks with ultra-long sequences. In addition, there are other high-performance LLMs such as LaMDA[32], GLM[33], OPT[34], and Chinchilla[35]. These LLMs have achieved inspiring results in natural language processing tasks, gradually revolutionizing the work patterns of professionals across various fields, including remote sensing[36], chemistry[5,37], medicine[38-40], and so on. In the field of geophysics, LLMs with geophysical knowledge can serve as the foundation for developing geophysical artificial intelligence, leveraging multimodal alignment to access various GeoFMs.

**2.2 Large Vision Models**

Most large vision models (LVMs) are primarily built upon Vision Transformer (ViT)[41] or convolutional neural network (CNN) architectures and are pretrained on large-scale image and video datasets. These models have achieved state-of-the-art results in various downstream tasks. In segmentation task, SAM[2] has attracted widespread attention due to its outstanding segmentation performance and remarkable generalization ability. Moreover, it can be used for the segmentation of data in other fields, such as picking the first arrivals in seismic data. Recently, an image restoration foundation model named SUPIR[42] was proposed. This model has achieved advanced results in various types of image restoration tasks. In addition, multimodal LVMs have made significant breakthroughs in text-to-image generation and text-controlled image editing, such as the contrastive language-image pre-training (CLIP) model[43], DALL-E[44] 2, Google Imagen[45], Stable Diffusion[46], etc.

The current LVMs have achieved excellent results in various visual tasks, but most of them are trained for single visual task or dependent on LLMs. Therefore, the development of pure large visual models has become a recent hotspot[47]. The success of LLMs is largely due to the in-context learning[26] method, which enables them to complete any text-formatted tasks through prompting. However, the absence of the in-context learning paradigm in LVMs hinders the flexible specification of language tasks in visual prompts as in LLMs[47]. Recently, researchers have started to explore new training paradigms for LVMs, with the objective of developing universal LVMs capable of tackling a variety of visual tasks. Bai et al. [48] (2023) defined a common format, "visual sentences", to enable the specification of visual tasks, and constructed a large-scale dataset to support their work. Specifically, "visual sentences" articulate visual tasks through sequences of images, with the model being capable of predicting the next image by learning patterns within the provided image sequence. Guo et al.[49] (2024) proposed a method of tokenizing the images first, followed by an autoregressive training on the tokens. These studies will advance the evolution of general LVMs, providing valuable insights for the progress of foundation models in other disciplines[50,51].

## 3   Foundation Models with Geophysics

Figure 2 illustrates the overview of GeoFMs in geophysics. Existing GeoFMs, trained on multimodal geophysical data, can accomplish various downstream tasks through interactions and are applicable to multiple subfields within geophysics. In this section, we review the development and applications of GeoFMs in the field of geophysics, including remote sensing, seismology, and other related sub-disciplines. Exploration geophysics is a focal point of this perspective, we will explore the applications and future directions of GeoFMs in exploration geophysics in the next section.

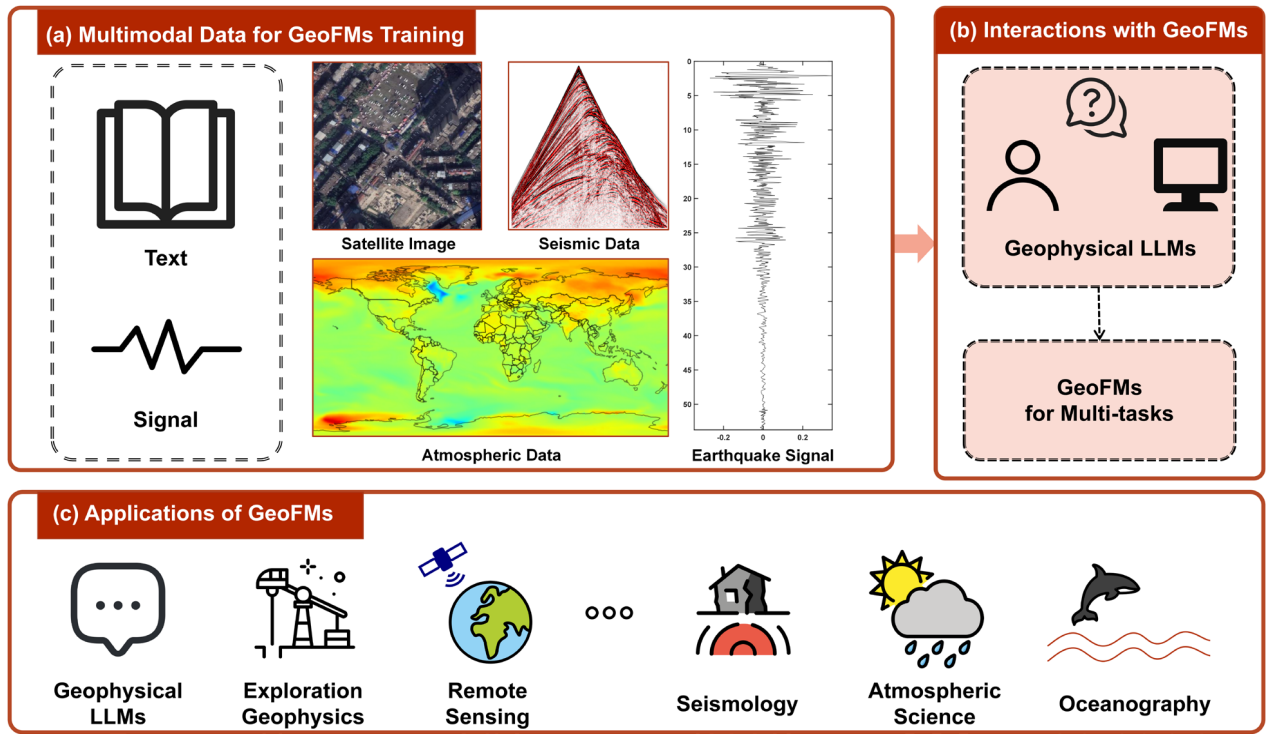

**Figure 2** Overview of GeoFMs in geophysics. (a) GeoFMs are trained on multimodal geophysical data to acquire knowledge in the geophysical domain. (b) Various tasks in the field of geophysics can be effectively accomplished by interacting with geophysical LLMs and invoking different GeoFMs. (c) Applications of GeoFMs in the field of geophysics.

## 3.1 Remote Sensing

Remote sensing, which employs sensors on satellites or aerial platforms, is a crucial tool for collecting geophysical data, facilitating the monitoring and analysis of the Earth's surface and environment. Remote sensing data primarily consists of various types of image data, including Synthetic Aperture Radar (SAR) images, thermal infrared images, optical images, among others. Currently, foundation models, trained on substantial image datasets, are capable of executing image processing tasks such as image segmentation and edge detection, which are also essential in the field of remote sensing. Given the analogous data types and tasks, fine-tuning LVMs with remote sensing data has achieved effective results, significantly propelling the progress of foundation models within the remote sensing domain.

The development of foundation models necessitates extensive training data. In recent years, numerous large-scale datasets in the field of remote sensing have been collected for training foundation models in this domain. The fMoW dataset[52], comprising over one million satellite images from over 200 countries, has been proposed for vision-related tasks in the field of remote sensing. The SatlasPretrain dataset[53] contains images with 302 million labels, offering a large-scale resource with a diversity of labels. The RSVG dataset[54] includes a substantial collection of image-query pairs, which can be used for the training of multimodal foundation models in remote sensing. These large-scale datasets provide a data foundation for the development of GeoFMs in the field of remote sensing.

Recently, a variety of GeoFMs have been introduced to address diverse tasks within the field of remote sensing. Kuckreja et al. [36] (2024) developed GeoChat by fine-tuning the LLaVA architecture[55] with a newly created remote sensing multimodal dataset, enabling the model to process high-resolution data and support interpretive human-machine interaction through natural language. Mall et al. (2024) [12] introduced ground remote alignment for training (GRAFT), a novel method for linking text with remote sensing images, facilitating the training of a comprehensive vision-language model without the need for costly language-paired data.

## 3.2 Seismology

Seismology is the scientific study of earthquakes and the propagation of elastic waves through the Earth or other planet-like bodies. This field involves analyzing and interpreting seismic data collected from seismographs and other instruments, thereby enhancing the understanding of the Earth's internal structure. Recent advancements in deep learning have significantly enhanced the field of seismology, with these methods now being broadly applied across a diverse range of tasks within the discipline. These tasks include but are not limited to phase picking[56-58], first-motion polarity classification[56,59], earthquake detection[58,60,61], event location[62,63], and earthquake prediction[64-66]. However, these methods are often trained on specific tasks using limited datasets, which restricts their wide application due to the limited generalization ability.

Inspired by the success of foundation models in various field, researchers have begun to explore the use of vast amounts of seismic data to train GeoFMs in the field of seismology. SeisCLIP[67] is a foundation model in the field of seismology, utilizing contrast learning on multimodal data for pretraining. This foundation model can be applied to a variety of downstream tasks such as event classification, localization, and focal mechanism analysis tasks through fine-tuning with small datasets. Li et al.[68] (2024) introduced Seismogram Transformer (SeisT), a foundation model designed for various earthquake monitoring tasks including earthquake detection, seismic phase picking, first-motion polarity classification, and so on. This model was trained on the DiTing dataset and evaluated for its generalization ability on the PNW dataset. Both datasets encompass a significant number of seismic events and corresponding labels such as arrival times, magnitude, and first-motion polarity. These GeoFMs have demonstrated substantial potential to address different seismological tasks, significantly impacting the research paradigm within the field of seismology.

### 3.3 Atmospheric Science

Atmospheric science is a scientific discipline dedicated to the study and understanding of the Earth's atmosphere, including aspects such as weather and climate and their interactions with other systems on Earth. This field is crucial for weather forecasting, climate prediction, and understanding changes in our environment due to natural and human-made factors. Over the past decade, weather forecasting has consistently relied on numerical weather prediction methods[69,70], which simulate transitions of atmospheric states based on partial differential equations.

Recently, deep learning-based weather prediction methods[71,72] have emerged as promising tools for accelerating weather forecasting. However, these methods, trained with limited data, do not achieve the accuracy of conventional numerical weather prediction techniques. To address these limitations, several GeoFMs have been developed in the field of atmospheric science. Trained with vast amounts of data, these GeoFMs have exhibited outstanding performance in both prediction speed and accuracy. Zhang et al.[73] (2023) developed NowcastNet using an end-to-end optimization architecture that integrates physical-evolution schemes to generate high-resolution, physically plausible nowcasts. Bi et al.[74] (2023) presented Pangu-Weather, a GeoFM designed for rapid and accurate weather prediction, which was trained on 39 years of global weather data. This model demonstrates superior performance compared to the leading numerical weather prediction model of that period, thereby highlighting the transformative potential of GeoFMs in atmospheric science.

### 3.4 Oceanography

Oceanography, or ocean science, investigates the intricacies of the oceans that cover over 70% of the Earth's surface. This field is vital for understanding marine life and biodiversity, assessing the ocean's role in climate regulation, and examining their impact on global economies. Inspired by the remarkable success of foundation models in general domain, oceanographer have begun to explore the potential of GeoFMs in the field of Oceanography. Bi et al.[75] (2024) introduced OceanGPT, the first oceanographic LLM pre-trained for various ocean science tasks. To overcome the challenges of acquiring ocean data, the domain construction framework named Doinstruct was proposed, enabling the construction of an ocean instruction dataset through multi-agent collaboration. Xiong et al.[76] (2023) presented AI-GOMS, a GeoFM employing the Fourier-based masked autoencoder architecture, which was designed for predicting ocean variables over a 30-day period.

# 4 Applications and Perspectives of Foundation Models in Exploration Geophysics

Unlike in areas such as medicine, remote sensing, and other geophysical disciplines, the application and development of GeoFMs in exploration geophysics are still in an initial stage. The development of GeoFMs in exploration geophysics face several challenges: 1) High reliance on physical principles: Many geophysical tasks depend on solving PDEs and adhering to physical laws. However, existing large-scale models have not demonstrated sufficient performance in these domains. 2) Data scarcity: The open-source exploration seismic data is limited in quantity and lacks corresponding labels. 3) Task complexity: Seismic data processing tasks are typically more complex compared to other fields such as image processing. Recently, Sheng et al.[77] (2023) proposed the first seismic foundation model, achieving state-of-the-art results in various seismic downstream tasks, which marks a significant exploration in geophysical research in the era of foundation models. In this section, we discuss the potential applications of large models in the procedures (seismic data processing, imaging, and interpretation) of exploration geophysics, and illustrate the immense potential of large models in the field of geophysics.

Figure 3 shows the overview of the applications of GeoFMs. For different seismic tasks, there will be various seismic foundation models to process data of different modalities to obtain the desired results. LLMs with exploration geophysics knowledge can used to interact with users and call different GeoFMs to complete various seismic tasks through multimodal alignment.

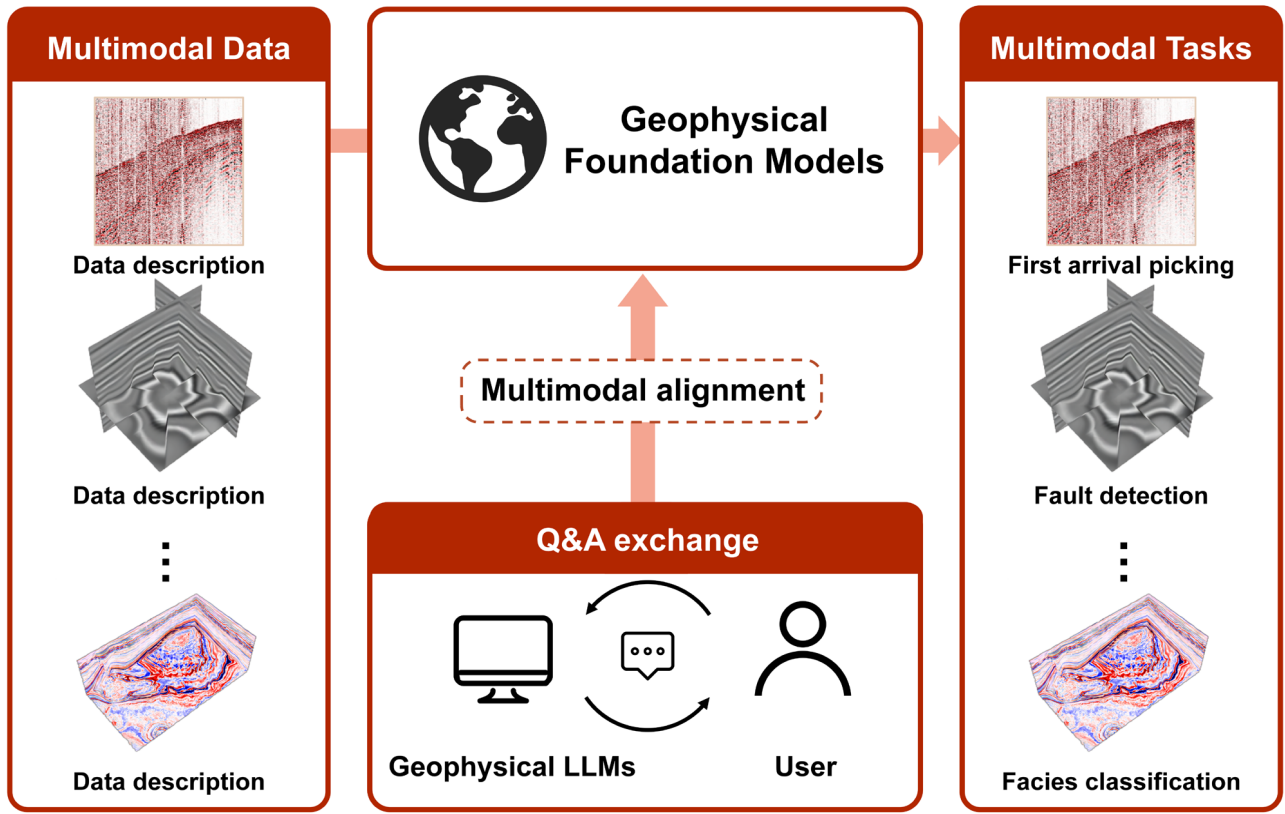

**Figure** 3 Overview of the GeoFMs application in the field of exploration geophysics. Users issue commands by interacting with seismic LLMs to process different modalities of input data through multimodal alignment using the suitable foundation model, and outputting the results of target tasks, such as first-arrival picking, fault detection, facies classification, etc.

### 4.1 Seismic Data Processing

Seismic data processing aims to transform the collected raw seismic records into a format that facilitates subsequent seismic imaging and interpretation. In this subsection, we present several potential applications for foundation models in geophysics, including first arrival picking based on SAM, as well as interpolation and denoising.

#### 4.1.1 First-arrival Picking Based on SAM

In seismic data processing, first-arrival picking plays a pivotal role in estimating the subsurface velocity structure, as it pinpoints the initiation of the first signal arrivals. In recent years, deep learning methods have been able to achieve excellent first-arrival picking results on similar datasets after training. However, the generalization performance of deep learning methods is significantly affected by the noise and differences in data distribution.

SAM[2], developed by Meta AI, is considered the first foundation model for computer vision. With its training conducted on a massive corpus of data, which encompasses millions of images and billions of masks, SAM demonstrates a robust capability of delivering effective segmentation results across a wide range of image segmentation tasks. The first-arrival picking in seismic data is also a segmentation task. Thus, employing SAM for this task presents promising prospects. Therefore, we propose the application of SAM directly to the task of first-arrival picking in seismic data, to demonstrate the impact of large models on the paradigm of exploration geophysics research.

SAM is composed of three parts: a prompt encoder, an image encoder, and a mask decoder. The prompt encoder serves to encode the prompt (mask, points, box, and text) into embedding, while the image encoder is a pre-trained model based on the ViT[41] architecture. The mask decoder is a lightweight module that updates both image and prompt embeddings through cross-attention, which is ultimately used for dynamic mask outputs[2]. The process of the first-arrival picking based on SAM is shown in Figure 4. There are two ways to use SAM. The first is to directly utilize the automatic segmentation feature of SAM and then select the largest mask as the first-arrival picking result, since the seismic data usually occupies the largest part of the image. The second method involves manually setting prompts, which can achieve better results when a more refined segmentation is needed, and it has a faster running speed. However, the drawback is that it requires manual settings of the prompts.

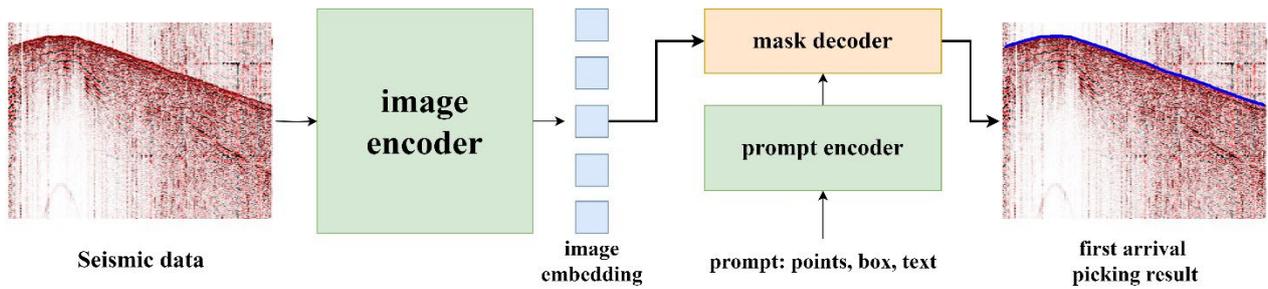

**Figure** 4 Overview of SAM-based first-arrival picking method. The seismic data and the prompt are encoded separately into embeddings by the image encoder and the prompt encoder, and subsequently fed into the mask decoder to generate the first-arrival picking results.

To better showcase the high generalization ability of SAM and its capability to fully automate the first-arrival picking process, we present the results obtained by the first method. The seismic data used for testing are derived from the public Halfmile and Sudbury datasets, published by St-Charles et al.[78] (2021). These datasets are very noisy, making it challenging to obtain accurate first-arrival picking results. The first-arrival picking labels for 84.45% of the traces were provided by experts in these two datasets. Figure 5a-c present the line gathers and corresponding labels from the Halfmile and Sudbury datasets, which are challenging for manual picking and only provide a limited number of labels across the three line gathers. To demonstrate the generalization ability of the SAM-based method on seismic data and its robustness to noise, we directly apply SAM to the data without any special treatment. And the results, represented by green points, are shown in Figure 5d-f. Even in areas of strong noise interference where no labels are provided, the SAM-based first-arrival picking method can still produce effective results.

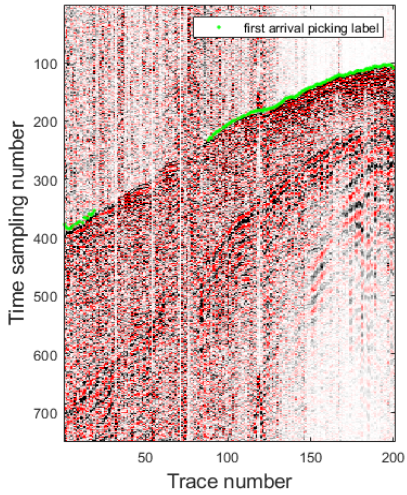 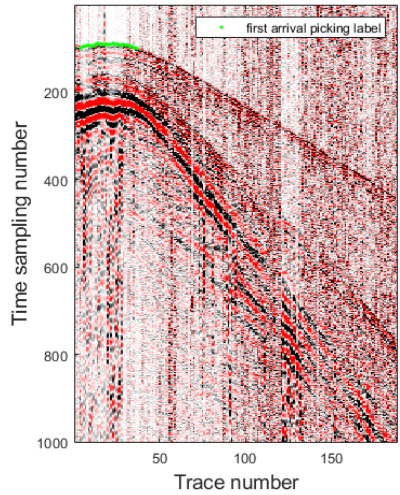 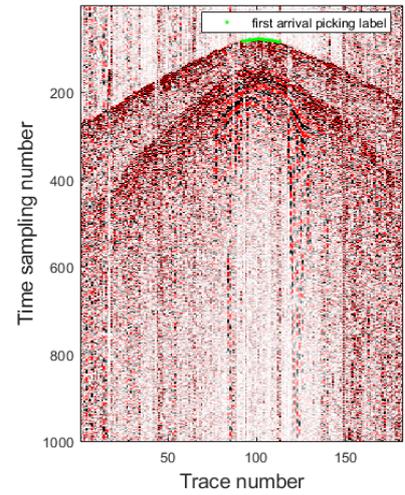

(a) test data (Halfmile)    (b) test data (Sudbury)    (c) test data (Sudbury)

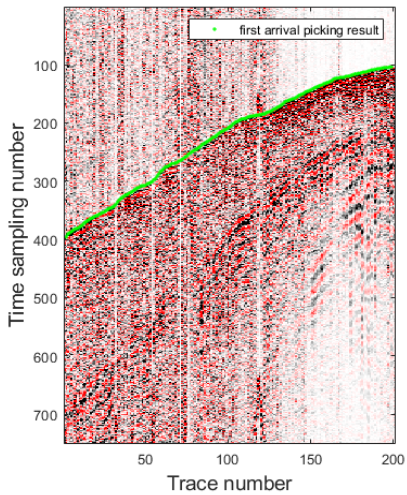 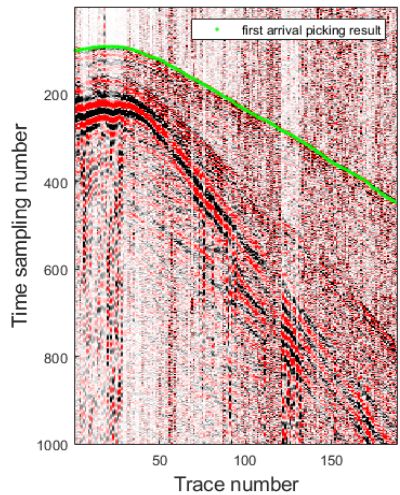 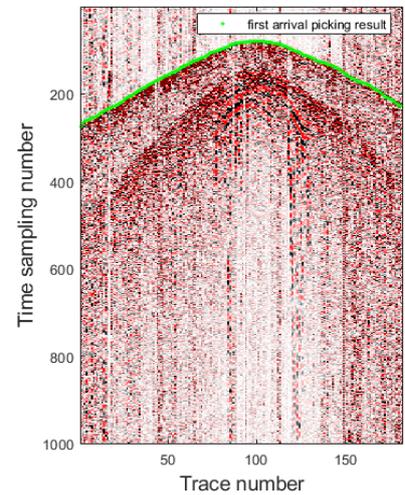

(d) SAM-based result (Halfmile)  (e) SAM-based result (Sudbury)  (f) SAM-based result (Sudbury)

**Figure** 5 SAM-based first arrival picking results. (a-c) Test data from the Halfmile, Sudbury datasets, and the provided first-arrival picking labels, which were deemed valid by experts. (d-f) First-arrival picking results based on SAM.

Despite the excellent performance of the SAM-based first-arrival picking method on seismic data, there are still many issues worthy of further study. First, the efficiency of the current SAM-based method is relatively low. Taking the Halfmile dataset as an example, it takes about 5 seconds to process one line gather without parallelization, which is far from industrial-level applications. Hence, accelerating the SAM-based first-arrival picking method is an important research direction in the near future. Secondly, the existing SAM has been trained on image and text data and has not been adapted to seismic data. Consequently, it is unable to execute semantic segmentation tasks on seismic data, such as fault detection, identifying areas containing specific noise based on provided prompts, and so on. Potential future directions involve utilizing technologies such as Adapter[79] and LoRA[80] to fine-tune SAM on seismic data. These methods have already made breakthroughs in fields such as medicine[81] and remote sensing[82]. The first-arrival picking method based on SAM demonstrates considerable potential and superior performance, which could potentially influence the research paradigm in this field.

### 4.1.2  Interpolation and Denoising

Due to the limitations of the data collection conditions, raw seismic records always have problems such as missing data or containing various types of noise. The purpose of seismic data interpolation is to recover data from sparse samples, while the aim of denoising is to separate valid signals from noise. In the practice of seismic data processing, seismic data are typically transformed into two-dimensional line gathers, similar to the format of images. The resemblance has led to the adoption of image processing techniques in most of the contemporary deep learning-based seismic data processing methods. Consequently, we can also adopt the image processing foundation models to develop seismic data processing foundation models.

In the raw seismic data, the missing data and noise are very complex, and the distribution of data varies wildly among different seismic data, impeding the generalization abilities of models and thus affecting the wide application of deep learning methods. To ensure that the trained model demonstrates robust generalization across various seismic datasets and processing tasks, there are two prevailing strategies for developing foundation models. The first strategy is based on pre-training and fine-tuning.

This strategy initially utilizes a large-scale dataset to pretrain a model with a large number of parameters based on the Mask Autoencoder[83] (MAE) framework, and then is fine-tuned in downstream tasks to achieve superior performance. Recently, Sheng et al.[77] (2023) developed the first seismic foundation model based on the pre-training and fine-tuning strategy. This model has demonstrated remarkable performance in seismic downstream tasks.

The second strategy is based on the seismic data priors of generative models, such as diffusion model[84,85], and generative adversarial network[86] (GAN). This strategy consists of two stages. The first stage is usually a corruption encoder, which encodes different types of missing data and noise into latent space. Subsequently, the second stage uses the embedding obtained from the first stage as prompts and employs a pre-trained diffusion model as the data prior for data restoration. Lin et al.[87] (2023) proposed diffusion models for the blind image restoration problem (DiffBIR), which pre-trained a restoration module to improve generalization capability and leveraged fixed stable diffusion through LAControNet for reconstruction. Wang et al.[88] (2023) presented StableSR, which only requires the fine-tuning of a lightweight, time-aware encoder to capture the degradation features.

In the realm of seismic data, the development of foundation models for interpolation and denoising presents substantial challenges. First, there is a notable lack of pre-trained models based on generative models serving as data priors in geophysics. Secondly, in contrast to image processing scenarios, seismic data embodies a wider variety and complexity of noise types. Consequently, the task of training an encoder to grasp the features of noise becomes a significant problem. These issues could be potential directions for future research in the advancement of GeoFMs.

**4.2 Seismic Imaging**

Seismic imaging is a technique that creates images of the subsurface structures by analyzing seismic waves. It is a challenging problem because traditional methods such as the full waveform inversion (FWI) or seismic tomography are highly dependent on the constraints of physics equations. Currently, deep learning-based seismic imaging methods can be primarily classified into two categories. The first is end-to-end learning, which directly learns the mapping from seismic data to the imaging domain[22,89]. This learning strategy can yield promising results on synthetic datasets that are similar to the training data. However, once there are changes in data distribution or model parameters, the application of the trained model becomes limited by its generalization capability. The second category employs deep learning as an auxiliary tool to assist in completing certain steps in traditional seismic imaging methods. Sun et al.[90] (2023) utilized a learned regularization as a constraint to optimize the FWI process. Ovcharenko et al.[91] (2019) extrapolated low-frequency from the respective high-frequency components of the seismic data to provide low-frequency information for FWI based on deep learning.

At present, the success of foundation models is mainly on applications that rely on experience and generative tasks. However, the comprehension and learning of physical laws for the solutions of partial differential equations (PDEs) are still in an initial stage. Seismic imaging, such as FWI, relies heavily on the forward modeling of wave equations, which is also the most time-consuming part of FWI. Therefore, it is extremely challenging to achieve a direct mapping from seismic data to images based on GeoFMs in the short term. Here, we propose two potential directions of development for foundation models in the field of seismic imaging. First, the acceleration of FWI is intrinsically linked to the efficient forward modeling of wave equations. Therefore, the initial research direction is to develop foundation models for solving PDEs. Ye et al.[92] (2024) introduced PDEformer, a foundation model for solving PDEs based on a graph transformer architecture. However, this model was limited to one-dimensional PDEs, so there is still a gap before it can be applied in seismic imaging. The second possible research direction is to integrate foundation models with traditional methods to overcome the inherent bottlenecks in conventional methods, such as the lack of low-frequency information. Most of these data preprocessing modules can be integrated within the foundation models for the interpolation and denoising section discussed in Section 4.1.2. Inspired by the GeoFM of weather forecasting[74], predicting the gradient changes during the FWI process could be a strategy to accelerate FWI. Changing the gradient during the FWI process, similar to weather forecasting tasks, involves predicting a trend over a certain number of time steps at specific grid points, and can be corrected based on the result at the current moment.

### 4.3 Interpretation

The purpose of seismic interpretation is to identify the geological information, including subsurface structure and properties, thereby locating the target spots underground. The process of seismic interpretation includes, but is not limited to, fault detection, geobody identification and segmentation, facies classification, and attribute analysis. The interpretation process is hindered by several limitations, such as the extensive data volume, inherent complexity, time-consuming nature, and uncertainties introduced by expert experience.

Since seismic interpretation involves processing and analyzing multimodal data, geophysical LLMs, which possess comprehensive knowledge in the field of geophysics, will serve as the foundation of multimodal GeoFMs. The data of different modalities are encoded by foundation models, and geophysical LLMs are used to interpret the instructions given by users and to invoke foundation models for different interpretation tasks. For the development of LLMs in geophysics, training data can come from various sources such as geophysics scholar publications, open-source codes, and programming interfaces. Deng et al.[93] (2023) introduced K2, the first open-source LLM for geoscience. K2 was trained on 5.5 billion tokens of geoscience text corpus, which were drawn from over a million of geoscience-related literature, based on the LLaMA-7B model. Lin et al.[94] (2023) developed GeoGalactica, a specialized open-source model derived from further pre-training of Galactica[95], with 30 billion parameters. This model has demonstrated state-of-the-art performance in diverse natural language processing tasks within the geoscience domain. Kuckreja et al.[36] (2024) presented GeoChat as an endeavor to expand the application of multimodal instruction-tuning[55,96] into the domain of remote sensing, with the objective of training a multitask conversational assistant. However, the exploration geophysics domain lacks a multimodal instruction-tuning conversational dataset, hence the establishment of large-scale training datasets for GeoFMs is one of the most important research directions in the future.

## 5 Future Directions and Challenges of Geophysics in the Era of Foundation Models

### 5.1 Future Directions for Developing GeoFMs

Currently, foundation models have achieved notable accomplishments across various domains, potentially influencing the scientific research paradigm in geophysics. Here, we present two potential directions for developing GeoFMs.

**1) Transferring based on existing foundation models:** Many tasks in geophysics, such as first arrival picking and image segmentation, fault detection, object detection, etc., are similar to those in the fields of computer vision and natural language processing. The existing large models have already achieved state-of-the-art results in tasks such as image segmentation, object detection, text-based language tasks, and others, with excellent generalization capabilities. Therefore, transferring the existing models to the field of geophysics is a promising strategy. Geophysics tasks suited for the transferring strategy share a common characteristic: the input data can be converted to an image format, and the output is a range rather than a specific numerical value. Consequently, there is no data precision reduction when seismic data are converted to image data. These tasks include first-arrival picking, fault detection, geobody identification and segmentation, facies classification, and human-computer interaction, among others. However, general foundation models have not been specifically trained with seismic data, and are devoid of basic geophysical concepts, thus inhibiting their ability to provide meaningful outputs. Consequently, it is necessary to fine-tune these foundation models on geophysical data when the outputs necessitate the understanding of geophysical concepts. The fine-tuning process introduces the domain knowledge of geophysics while preserving the superior performance of the existing foundation models. Compared to training foundation models from scratch, this transfer strategy does not require massive amounts of training data but demands high quality data.

**2) Training GeoFMs from Scratch:** In the field of geophysics, most tasks are more difficult compared to those in the computer vision or natural language processing fields. For instance, in seismic data processing, the types of data corruption and noise are more complex than in image processing tasks, and seismic data are stored as floating-point numbers while image data are stored as integers. This makes it difficult for the existing foundation models to be transferred to the field of seismic data processing, including denoising, interpolation, and seismic imaging. Therefore, training from scratch is a necessary strategy to develop GeoFMs, and potential avenues for developing GeoFMs for these tasks have been discussed in Section 4.

## 5.2 Challenges

Despite the promising prospects of GeoFMs, there are still challenges in the development and application of GeoFMs in the field of geophysics.

**1) Data:** Although the volume of open-source geophysical data is massive, reaching TB scale, there are several limitations for the development of GeoFMs. (1) Insufficient data diversity: While a single exploration area may have a large volume of seismic data, there is a significant difference in data distribution among different exploration regions. To enable the GeoFMs to have better generalization ability, diverse data from different exploration regions are required, rather than large amounts of data from a few target areas. (2) Lack of labels: For field seismic data, there is a shortage of corresponding label data for GeoFMs training. (3) Access restrictions: Due to confidentiality as well as legal and privacy issues, some geophysical data are not permitted to be used for training and publication.

**2) Benchmark:** Deep learning methods have achieved great success and development in the field of exploration geophysics. However, there is currently a lack of unified benchmarks for comparison in the field. Different researchers employ different seismic data preprocessing procedures, which may result in discrepancies in the results obtained even when using the same neural network for the same task. This makes it difficult to judge the merits of different models in this field under a unified standard.

**3) Computation:** Current large models, whether for training or fine-tuning, consume a substantial amount of time and resources which are beyond the reach of most researchers and institutions. Even merely deploying and utilizing large models requires a significant number of computational resources. This has led to a situation where access to cutting-edge foundation model technology is often limited to well-funded organizations, creating a barrier for smaller institutions and independent researchers.

# 6 Conclusions

In this perspective, we introduce the current state of foundation models and their potential impact on the paradigm of geophysical research, including as follows:

1. In the era of foundation models, the focus of researchers is shifting from traditional methods that rely on prior assumptions to the development of large-scale GeoFMs, and many GeoFMs have been introduced to the field of geophysics. GeoFMs have shown enormous potential for addressing geophysical tasks such as data processing, data interpretation, and tasks that traditionally depend on empirical knowledge and expert intervention. Thus, GeoFMs can partially replace part of traditional methods and play a significant role in geophysical applications.

2. Existing foundation models face challenges in addressing problems that are highly constrained by physical principles, such as inversion imaging in exploration seismology. A critical area for future research will involve strategies to integrate complex physical theories into GeoFMs, or to combine GeoFMs effectively with conventional methods to advance geophysical imaging techniques, potentially enhancing the efficiency of geophysical imaging fundamentally.

3. The development of GeoFMs is still in an initial stage. This perspective outlines two development strategies for GeoFMs. The first strategy involves using transfer learning methods for geophysical tasks similar to those in computer vision or natural language processing. This method entails fine-tuning existing large models with geophysical data to accomplish tasks in the geophysical domain. The second strategy involves collecting a vast array of geophysical data tailored for different tasks and training GeoFMs from scratch to tackle various geophysical issues.

By training on specific geophysical tasks and learning domain knowledge, GeoFMs are capable of handling multi-modal data, thus making it applicable to almost all seismic tasks. Given that the development of GeoFMs is still in its initial exploratory stages, we propose two strategies for the advancement of GeoFMs, and present the current state and training methods used in similar domains for foundation models associated with various tasks. Despite the promising prospects of GeoFMs, it will also face many challenges, including the lack of training data, privacy issues, lack of computational resources, among others. The development of GeoFMs will provide limitless possibilities for the advancement of technology in the field of geophysics and will also transform the research paradigm in this field. The journey ahead may be fraught with challenges, but with continuous exploration and innovation, we are poised to witness a revolution in the field of exploration geophysics.